\documentclass[aps,prd,amsmath,amssymb,showpacs]{revtex4}

\usepackage{graphicx}
\begin{document}

\title{Top quark spin polarization  in ep collision}

\author{S. Ata\u{g} }
\email[]{atag@science.ankara.edu.tr}
\affiliation{Department of Physics, Faculty of Sciences, 
Ankara University, 06100 Tandogan, Ankara, Turkey}

\author{B. D. \c{S}ahin }
\email[]{dilec@science.ankara.edu.tr}
\affiliation{Department of Physics, Faculty of Sciences,
Ankara University, 06100 Tandogan, Ankara, Turkey}

\begin{abstract}
We discuss the degree of spin polarization of single top quarks 
produced via $Wg$ fusion process in $ep$ collision at TESLA+HERAp
and CLIC+LHC energies $\sqrt{s}=1.6$ and 5.3 TeV.
For $eb \to t \bar{\nu}$ subprocess we show that
the top quark spin is completely polarized
when the spin basis is chosen in the direction of
the incoming positron beam in the rest frame of top quark.
A description on how to combine the cross sections 
of $e^{+}b\to t\bar{\nu}$ and $e^{+}g\to t\bar{b}\bar{\nu}$ processes 
is given. $e^{+}$-beam direction is  taken to be
 the favorite  top quark spin decomposition axis  
in its rest frame and it is found to be comparable with 
the ones in $pp$ collision. 
It is argued that theoretical simplicity and experimental 
clearness are the advantage of  $ep$ collision. 

(Revised version of Phys. Rev. {\bf D69}(2004)034016.)
\end{abstract}


\maketitle

\section{Introduction}
The large mass of the top quark implies that its weak decay 
time is much shorter than the typical time for the  strong 
interactions to affect its spin \cite{zerwas}. 
Within the standard model, the dominant decay chain of the 
top quark is $t\to W^{+}b\, (W\to \ell^{+}\nu,\,\, \bar{d}u )$. 
The  angular distribution of the top  decay  
given below is a simple expression 
to clarify the correlations 
between the top quark decay products and the top spin:
\begin{eqnarray}
\frac{1}{\Gamma_{T}}\frac{d\Gamma}{d\cos \theta}=
\frac{1}{2}(1+\alpha \cos \theta)
\end{eqnarray}
Here  $\theta$ is defined as the angle between top quark
decay products and top quark spin quantization axis in the 
rest frame of the top quark. 
For $\ell$ or $\bar{d}$ the correlation coeficient 
$\alpha=1$ which leads to the strongest correlation. 
Therefore, it is expected that the study of 
the single production of top quark will give a useful tool
to understand the standard model mechanism for electroweak 
symmetry and coupling of the top quark to other particles.
There are many detailed discussions in the literature for 
the single top production and spin correlations
in $pp$ and $p\bar{p}$ collisions \cite{mahlon,parke}.

The purpose of this work is to describe polarization of the 
top quarks along the direction of various spin bases for the 
single production in $ep$ collision via $e^{+}b\to t\bar{\nu}$ 
and $e^{+}g\to t\bar{b}\bar{\nu}$ processes. 

It is hoped that the linear $e^{+}e^{-}$ collider can be  
converted into an $ep$ collider as an additional option 
when linear collider is constructed on the same base as the 
proton ring. Linear collider design TESLA at DESY is the 
one which has this option called TESLA+HERAp $ep$ collider
\cite{tesla}.
Similar option would be considered for CLIC+LHC at CERN.
We will take into account the estimations about the energy 
and luminosity of these colliders as $\sqrt{s_{ep}}=1.6$ 
TeV ($E_{p}=800, E_{e}=800$ GeV),
$L_{ep}=10^{31}$ $cm^{-2}s^{-1}$ for TESLA+HERAp
and $\sqrt{s_{ep}}=5.3$ TeV ($E_{p}=7000, E_{e}=1000$ GeV) ,
$L_{ep}=10^{32}$ $cm^{-2}s^{-1}$ for CLIC+LHC.

At $pp(\bar{p})$ colliders top quark decay products occur 
in both the $t\bar{t}$ and the single $t$ production modes.
In the case of pair production, there is potentially 
observable angular correlations among the decay products 
arising from the fact that spin up top quark is more likely 
to be produced along with a spin down top antiquark. The 
size of these 2-particle correlations tends to be smaller 
in $gg$ collisions than in $q\bar{q}$ collisons.
Because the dominant contribution to top quark pair production 
is due to the $gg$ initial state, the predicted 2-particle 
correlations are rather small \cite{mahlon2}. 
The single top quark production has the  electroweak process
with produced top quarks couples to a $W$ boson. That is why 
we expect strong spin polarization of top quarks. But single 
top decay products in the final states give smaller statistics 
when compared to pair production. Therefore, a sophisticated  
signal analysis needs to be performed 
at $pp(\bar{p})$ colliders.
 
In $ep$ collision top quark decay products in the 
final states are dominated by the single $t$ 
production due to  absence of the $t\bar{t}$ production.
This is a great advantage that the single top 
production is a result of primarily electroweak 
process. Therefore, it is anticipated that the $ep$ 
colliders with linear electron (or positron) beams are 
appropriate choice  for investigating top quark spin 
polarization.

\section{Cross Sections for Single Production of 
Polarized Top Quark}

The calculation of single top quark production cross section 
is based on the $Wg$ fusion process with 
$e^{+}g\to t\bar{b} \bar\nu$ 
diagram. These diagrams are dominated by the configuration 
where the final state $\bar{b}$ quark is nearly collinear with 
the incoming gluon. If b quark mass is neglected the cross section
becomes singular. Furthermore, at each order in the strong coupling,
there are logarithmically enhanced contributions, converting the 
perturbation expansion from a series in $\alpha_{s}$ to one in 
$\alpha_{s}(\mu^{2})\ln{(\mu^{2}/m_{b}^{2})}$. Here  
$\mu^{2}=Q^{2}+m_{t}^{2}$ and $Q^{2}$ is the virtuality of the 
$W$ boson.
The slow convergence of this original perturbation expansion 
caused by the large logarithms can be absorbed into the 
b quark distribution function by  resumming the terms  
to all orders. Once the b quark parton distribution function 
has been introduced \cite{olness}, 
one should begin with $2\to 2$ process 
such as $e^{+}b\to t\bar{\nu}$. The $2\to 3$ process still 
can be included  as a correction to the $2\to 2$ process.
Because the logarithmic part of the $2\to 3$ process is 
already covered by b quark parton distribution function 
used to compute $2\to 2$ process, it is necessary to subtract 
the overlap region of these two diagrams to avoid double 
counting\cite{stelzer}. So, combined cross section becomes 

\begin{eqnarray}
[\sigma(eb\to t\bar{\nu})+\sigma(eg\to t\bar{b}\bar{\nu})
-\sigma(g\to b\bar{b}*eb\to t\bar{\nu})]
\end{eqnarray}
where the subtracted term is the gluon splitting piece of the 
cross section for $eg\to t\bar{b}\bar{\nu}$.

For the main process $e^{+}b\to t\bar{\nu}$ spin dependent squared 
amplitude is given by

\begin{eqnarray}
|M|^{2}=\frac{2g_{w}^{4}}{(q^{2}-M_{W}^{2})^{2}}(p_{b}\cdot p_{\nu})
[p_{e}\cdot p_{t}-m_{t}\, p_{e}\cdot s_{t}]
\end{eqnarray}
where momentum of the each particle is represented by its symbol. 
The momentum transfer q and top quark spin vector are 
defined by
\begin{eqnarray}
&&q=p_{b}-p_{t} \nonumber \\
&&s_{t}^{\mu}=(\frac{\vec{p}_{t}\cdot \vec{s^{\prime}}}{m_{t}}
\,,\, \vec{s^{\prime}}+\frac{\vec{p}_{t}\cdot \vec{s^{\prime}}
}{m_{t}(E_{t}+m_{t})}\vec{p}_{t}) \nonumber \\
&&(s_{t}^{\mu})_{R.S.}=(0, \vec{s^{\prime}})
\end{eqnarray}
with R.S. stands for rest system of the top quark.

In the case of $2\to 2$ process we will show that, the top 
quark spin is completely polarized along the  positron 
beam direction (spin up)  in top rest frame. 
The definition of the spin axis in the 
rest frame of the top quark does not depend on the 
coordinate system where the cross section is performed.
So it is more convenient to calculate cross section 
in the center of mass system of incoming positron and b quark.
The momentum definitions of incoming and outgoing fermions are
\begin{eqnarray}
&&p_{b}=\frac{\sqrt{s}}{2}(1, 0, 0, 1) \,\,\,\,
p_{e}=\frac{\sqrt{s}}{2}(1, 0, 0, -1) \\
&&p_{t}=(E_{t}, |\vec{p}_{t}|\sin{\theta}, 0,
 |\vec{p}_{t}|\cos{\theta})\\
&&p_{\nu}=(|p_{t}|, -|\vec{p}_{t}|\sin{\theta}, 0,
 -|\vec{p}_{t}|\cos{\theta})
\end{eqnarray}

Here Mandelstam variable $s$ belongs to subprocess 
of $eb$ scattering.
Top quark energy $E_{t}$ and momentum 
$|\vec{p}_{t}|$ can be written in terms 
of  $s$ and top mass $m_{t}$

\begin{eqnarray}
|\vec{p}_{t}|=\frac{s-m_{t}^{2}}{2\sqrt{s}}, \,\,\,\,\,\,
E_{t}=\frac{s+m_{t}^{2}}{2\sqrt{s}}
\end{eqnarray}

Let us define top spin direction along the positron beam 
as follows
\begin{eqnarray}
\vec{{s}^{\prime}}=\lambda \frac{\vec{{p}_{e}^{\star}}}
{|\vec{{p}_{e}^{\star}}|} ,\,\,\,\,\, \lambda=\pm 1. 
\end{eqnarray}
where $\vec{{p}_{e}^{\star}}$ is the positron momentum observed 
in the rest frame of the top quark. Since positron momentum 
$\vec{p}_{e}$  is first defined in CM system where 
the cross section is calculated, one should apply Lorentz boost 
to the rest frame of the top quark with the expression below

\begin{eqnarray}
\vec{{p}_{e}^{\star}}=\vec{p}_{e}+\frac{\gamma-1}{\beta^{2}}
(\vec{\beta}\cdot \vec{p}_{e})\vec{\beta}
-E_{e}\gamma \vec{\beta}
\end{eqnarray}
where $\vec{\beta}$ is the velocity of the top quark in the 
 CM system and $\beta^{2}$ can be written in terms 
of $s$ and $m_{t}$

\begin{eqnarray}
\vec{\beta}=\frac{\vec{p}_{t}}{E_{t}}, \,\,\,\,\,\, 
\beta^{2}=\frac{(s-m_{t}^{2})^{2}}{(s+m_{t}^{2})^{2}}, 
\,\,\,\,\,\, 
\gamma=\frac{1}{\sqrt{1-\beta^{2}}}
\end{eqnarray}

Using the above expressions we obtain scalar products of 
four vectors $p_{e}\cdot p_{t}$ and $p_{e}\cdot s_{t}$
inside the square bracket of the squared amplitude 

\begin{eqnarray}
p_{e}\cdot p_{t}=\frac{1}{4}[(s-m_{t}^{2}) \cos{\theta} + 
s+m_{t}^{2}] \\
p_{e}\cdot s_{t}=-\frac{\lambda}{4 m_{t}}[(s-m_{t}^{2}) 
\cos{\theta} + s+m_{t}^{2}]
\end{eqnarray}
When these results are put into the squared amplitude of 
process $e^{+}b\to t\bar{\nu}$ it is clear that the amplitude 
gives zero for $\lambda=-1$ (spin down). Therefore top spin 
is 100\% polarized when the spin direction  is 
chosen along the incoming e-beam (spin up) 
in the top quark rest frame.

For the correction to $2\to 2$ process $Wg$ fusion process 
$e^{+}g\to t\bar{b}\bar{\nu}$ we write the amplitude as a sum of 
the contribution from each diagram:

\begin{eqnarray}
|M|^{2}=|M_{1}|^{2}+|M_{2}|^{2}+|M_{int}|^{2}\\
|M_{1}|^{2}=\frac{g_{W}^{4}g_{s}^{2}}{(q_{1}^{2}-M_{W}^{2})^{2}
(q_{2}^{2}-m_{b}^{2})^{2}}\,\, A_{1}\\
|M_{2}|^{2}=\frac{g_{W}^{4}g_{s}^{2}}{(q_{1}^{2}-M_{W}^{2})^{2}
(q_{3}^{2}-m_{t}^{2})^{2}}\,\, A_{2} \\
|M_{int}|^{2}=\frac{g_{W}^{4}g_{s}^{2}}{(q_{1}^{2}-M_{W}^{2})^{2}
(q_{2}^{2}-m_{b}^{2})(q_{3}^{2}-m_{t}^{2})}\,\,A_{int} \\
\end{eqnarray}
where $M_{1}$, $M_{2}$ and $M_{int}$ correspond diagrams with
$b$ quark exchange, $\bar{t}$ quark exchange and their
interference, respectively. Reduced amplitudes $A_{1}$,
$A_{2}$ and $A_{int}$ are

\begin{eqnarray}
A_{1}=&&(p_{\nu}\cdot p_{b})(-p_{b}\cdot p_{g}\, p_{t}\cdot p_{e}+
m_{t}\,p_{b}\cdot p_{g}\, p_{e}\cdot s_{t}-
m_{b}^{2}\,p_{t}\cdot p_{e}+m_{t}\,m_{b}^{2}\,p_{e}\cdot s_{t})
\nonumber \\
&&+(p_{\nu}\cdot p_{g})(p_{b}\cdot p_{g}\,p_{t}\cdot p_{e}
-m_{t}\,p_{b}\cdot p_{g}\, p_{e}\cdot s_{t} 
+m_{b}^{2}\,p_{t}\cdot p_{e}
-m_{t}\,m_{b}^{2}\,p_{e}\cdot s_{t})
\end{eqnarray}

\begin{eqnarray}
A_{2}=&&(p_{\nu}\cdot p_{b})(-p_{t}\cdot p_{e}\,p_{t}\cdot p_{g}
+m_{t}\,p_{t}\cdot p_{e}\,p_{g}\cdot s_{t}-m_{t}^{2}\,p_{t}\cdot p_{e}
+p_{t}\cdot p_{g}\,p_{e}\cdot p_{g} \nonumber \\
&&-m_{t}\, p_{e}\cdot p_{g}\,
p_{g}\cdot s_{t}+m_{t}^{2}\, p_{e}\cdot p_{g}+
m_{t}^{3} \,p_{e}\cdot s_{t})
\end{eqnarray}

\begin{eqnarray}
A_{int}=&&(p_{\nu}\cdot p_{b})(2 \, p_{b}\cdot p_{t}\, p_{t}\cdot p_{e}
-p_{b}\cdot p_{t}\, p_{e}\cdot p_{g}- 2\,m_{t}\, p_{b}\cdot p_{t}\, 
p_{e}\cdot s_{t} \nonumber \\
&&+p_{b}\cdot p_{e}\, p_{t}\cdot p_{g}-m_{t}\, p_{b}\cdot p_{e}\,
p_{g}\cdot s_{t}+m_{t}\, p_{b}\cdot s_{t}\, p_{e}\cdot p_{g})
\nonumber \\
&&+(p_{\nu}\cdot p_{t})(p_{b}\cdot p_{g}\, p_{t}\cdot p_{e}
-m_{t}\, p_{b}\cdot p_{g}\,p_{e}\cdot s_{t}) \nonumber \\
&&+(p_{\nu}\cdot p_{g})(-p_{b}\cdot p_{t}\, p_{t}\cdot p_{e}
+m_{t}\, p_{b}\cdot p_{t}\, p_{e}\cdot s_{t})
\end{eqnarray}

\begin{eqnarray}
q_{1}=p_{e}-p_{\nu}\,\, , \,\,\,\, q_{2}=p_{g}-p_{b}\,\, , \,\,\,\,
q_{3}=p_{t}-p_{g}\,\, .
\end{eqnarray}

Gluon splitting part of the cross section that we should subtract
can be written as follows 

\begin{eqnarray}
\sigma(g\to b\bar{b}*eb\to t\bar{\nu})=\int_{m_{t}^{2}/s}^{1}
\hat{\sigma}(eb\to t\bar{\nu})f_{b/p}^{LO}(x,\mu^{2})dx
\end{eqnarray}
where $f_{b/p}^{LO}$ is the probability for a gluon to split 
into $b\bar{b}$ pair at leading order

\begin{eqnarray}
f_{b/p}^{LO}(x,\mu^{2})=\frac{\alpha_{s}(\mu^{2})}{2\pi}
\ln{(\mu^{2}/m_{b}^{2})}\int_{x}^{1}\frac{dz}{z}P_{b/g}(z)
f_{g/p}(x/z , \mu^{2})
\end{eqnarray}
with splitting function

\begin{eqnarray}
P_{b/g}(z)=\frac{1}{2}[z^2+(1-z)^2].
\end{eqnarray}

In splitting procedure
the same QCD factorization scale should be considered
as the one used in the parton distribution functions.
All of the cross sections presented in this paper have been 
computed using MRST2002 parton distribution functions
\cite{stirling} and 
running $\alpha_{s}$ \cite{pdg}. Considering combination, we obtain 
unpolarized total cross section of single top quark 
production 3.2 pb at TESLA+HERAp and 32.5 pb at CLIC+LHC.

\section{Results and Discussion}

Let us now discuss the spin of the top quarks  produced 
with $e^{+}b\to t\bar{\nu}$ process ($2\to 2$ contribution) 
at TESLA+HERAp and CLIC+LHC energies $\sqrt{s}=1.6$ and 5.3 TeV. 
In the zero momentum frame(ZMF) of the initial state
partons t quark and antineutrino go out  back to back. 
Since they couple to a $W$ boson, in the
initial state b quarke has left handed chirality and positron 
right handed chirality. Their chiralities imply left and right 
handed helicities due to their high speed. Because of the angular 
momentum conservation outgoing t quark has both left and right 
handed helicity component. In this frame, it is shown from 
Table~\ref{tab1} and Table~\ref{tab2} that t quark 
is left handed 95\% and 97\% of the time for TESLA+HERAp 
and CLIC+LHC, respectively.
Helicity of the particle with 
large mass is highly frame dependent because the speed of t 
quark is not ultrarelativistic at the energy region of 
$ep$ colliders. So, in the laboratory frame of TESLA+HERAp 
and CLIC+LHC it is right handed 80\% and  62\% of the time.

In the case of $e^{+}g\to t\bar{b}\bar{\nu}$ process ($2\to 3$)
additional third particle in the final state shares 
the momentum, then the fraction of 
left handed helicity component decreases in the ZMF which 
is left handed only 76\%(77\%)  of the time
at TESLA+HERAp (CLIC+LHC) energies. It is right handed 
77\% and 62\% of the time in the laboratory(LAB) frame of 
TESLA+HERAp and CLIC+LHC. 

It should be pointed  out that the overlap region needs 
an extra discussion. There are two possibilities
to define ZMF which are in terms of either $eg$ or $eb$ 
initial states.
As explained above helicity of the top quark is not 
invariant under longitudinal boosts connecting these 
two frames. Therefore, it is not possible to define
ZMF uniqely.  A further discussion 
concerning its experimental side can be found in 
Ref.~\cite{mahlon} for $pp$ collision.

For a spin basis whose definition does not depend on 
the existence of a ZMF, 
we would think of  antineutrino direction as the 
decomposition axis of the top quark spin in theoretical 
point of view. Since the spin 
direction is defined in the rest frame of the top quark, 
choosing antineutrino axis is almost equivalent to 
observing the helicity of the t quark in the frame where 
antineutrino and t quark are back to back for 
$2\to 2$ process. In this case 
we obtain spin up contribution with 95\% of the time  
which has the same  degree of polarization as the  ZMF helicity 
basis for TESLA+HERAp.
At CLIC+LHC energy, fractions with left handed helicity 
in ZMF and with spin up in antineutrino direction are both 
97\% . 
In the $2\to 3$ case 90\% of top quarks are produced with 
spin up in this basis for TESLA+HERAp and 89\% for 
CLIC+LHC. Combination of $2\to 2$ and  $2\to 3$
processes with subtraction gives 92\%(91\%) fraction of spin up 
quarks at TESLA+HERAp (CLIC+LHC) energy.
In experimental point of view, the antineutrino is invisible
and we need somthing visible to make the basis definition.
Fortunately, such an alternate definition is possible. The 
outgoing antineutrino is only  slightly deflected from 
the incoming positron direction  in $2\to 2$ or $2\to 3$ 
process, and so it is interesting   to consider 
$e^{+}-beam$ line to define the spin basis. 
Then we find overall fraction of the spin up top quarks 
as 96\% which is close to the one in the antineutrino axis
at TESLA+HERAp collider. This becomes 93\% at CLIC+LHC collider.
In tables, we have kept results from the antineutrino 
direction for theoretical motivation.
From the rest frame of the top quark e-beam and p-beam are not
back to back. Then, it is also interesting to 
discuss the proton-beam (p-beam) as a spin decomposition 
axis for comparison. The last lines in the 
tables represent the results of p-beam axis which show
the spin down most of the time as expected.

Another quantity concerning the spin-induced angular correlations
is the spin asymmetry 
\begin{eqnarray}
A_{\uparrow\downarrow}=\frac{N_{\uparrow}-N_{\downarrow}}
{N_{\uparrow}+N_{\downarrow}} 
\end{eqnarray}
In the case where spin-up and spin-down top quarks are 
both present spin asymmetry appears in the differential 
distribution of the decay angle presented in the first section

\begin{eqnarray}
\frac{1}{\Gamma_{T}}\frac{d\Gamma}{d\cos \theta}=
\frac{1}{2}(1+A_{\uparrow\downarrow}\alpha \cos \theta)
\end{eqnarray}
The last column of the Table~\ref{tab1}  
shows that antineutrino 
or  $e^{+}$-beam basis improves the asymmetry a factor of 1.7
when compared to LAB helicity system.
From Table~\ref{tab2} we see that this factor takes the 
value of 4.  Clearly, it is easier to observe 
angular correlation as the asymmetry increases.
Comparison of Table~\ref{tab1} and Table~\ref{tab2} provides
that the fraction of top quarks produced with right 
handed helicity in LAB helicity frame gets smaller values 
as the center of mass energy of $ep$ system gets larger.

\begin{table}
\caption{Dominant spin fractions and asymmetries for the 
various top quark  spin bases in the production 
of single top  process with $Wg$ fusion channel at 
$\sqrt{s}=1.6$ TeV TESLA+HERAp energy. Contributions from 
each $2\to 2$  and $2\to 3$ processes  and 
combination of them are listed.
\label{tab1}}
\begin{ruledtabular}
\begin{tabular}{cccccc}
basis & 2$\to$2 & 2$\to$3 & overlap & total & 
$\frac{N_{\uparrow}-N_{\downarrow}}{N_{\uparrow}+N_{\downarrow}} $ \\
\hline
LAB helicity & 80\%(R) &77\%(R) &79\%(R) &77\%(R) &0.54 \\
ZMF helicity & 95\%(L) &76\%(L) &undefined&undefined &undefined \\
e-beam &100\%$\uparrow$  &93\%$\uparrow$ &98\%$\uparrow$ &
96\%$\uparrow$ &0.92 \\
Antineutrino & 95\%$\uparrow$ &90\%$\uparrow$ &93\%$\uparrow$ &
92\%$\uparrow$&0.84 \\
p-beam & 90\%$\downarrow$ &86\%$\downarrow$ &88\%$\downarrow$ &
87\%$\downarrow$&-0.75 \\
\end{tabular}
\end{ruledtabular}
\end{table}

\begin{table}
\caption{Dominant spin fractions and asymmetries for the
various top quark  spin bases in the production
of single top  process with $Wg$ fusion channel at
$\sqrt{s}=5.3$ TeV CLIC+LHC energy. Contributions from
each $2\to 2$  and $2\to 3$ processes  and
combination of them are listed.
\label{tab2}}
\begin{ruledtabular}
\begin{tabular}{cccccc}
basis & 2$\to$2 & 2$\to$3 & overlap & total &
$\frac{N_{\uparrow}-N_{\downarrow}}{N_{\uparrow}+N_{\downarrow}} $ \\
\hline
LAB helicity & 62\%(R) &62\%(R) &62\%(R) &60\%(R) &0.20 \\
ZMF helicity & 97\%(L) &77\%(L) &undefined&undefined &undefined \\
e-beam &100\%$\uparrow$  &91\%$\uparrow$ &96\%$\uparrow$ &
93\%$\uparrow$ &0.86 \\
Antineutrino & 97\%$\uparrow$ &89\%$\uparrow$ &94\%$\uparrow$ &
91\%$\uparrow$&0.83 \\
p-beam & 86\%$\downarrow$ &81\%$\downarrow$ &84\%$\downarrow$ &
82\%$\downarrow$&-0.65 \\
\end{tabular}
\end{ruledtabular}
\end{table}

\begin{figure}
\includegraphics{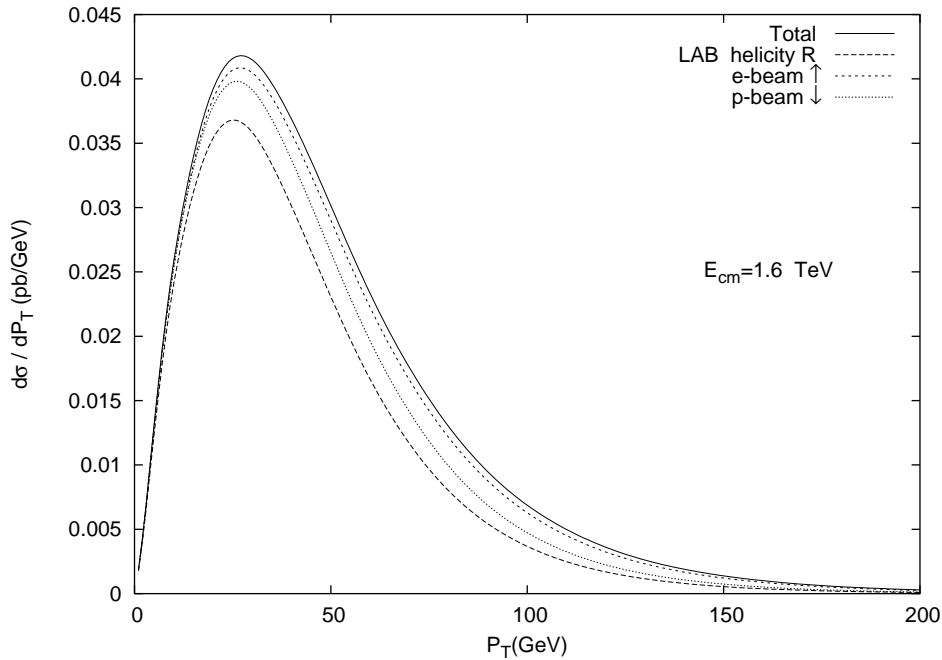}
\caption{Transverse momentum $P_{T}$ distributions
of the singly produced top quarks via $Wg$ fusion at 
TESLA+HERAp enegy $\sqrt{s}=1.6$ TeV. Dominant spin bases 
LAB helicity right, e-beam up, antineutrino up
and unpolarized(Total) case are drawn. Combination of 
$2\to 2$  and $2\to 3$ processes are considered.
\label{fig1}}
\end{figure}

\begin{figure}
\includegraphics{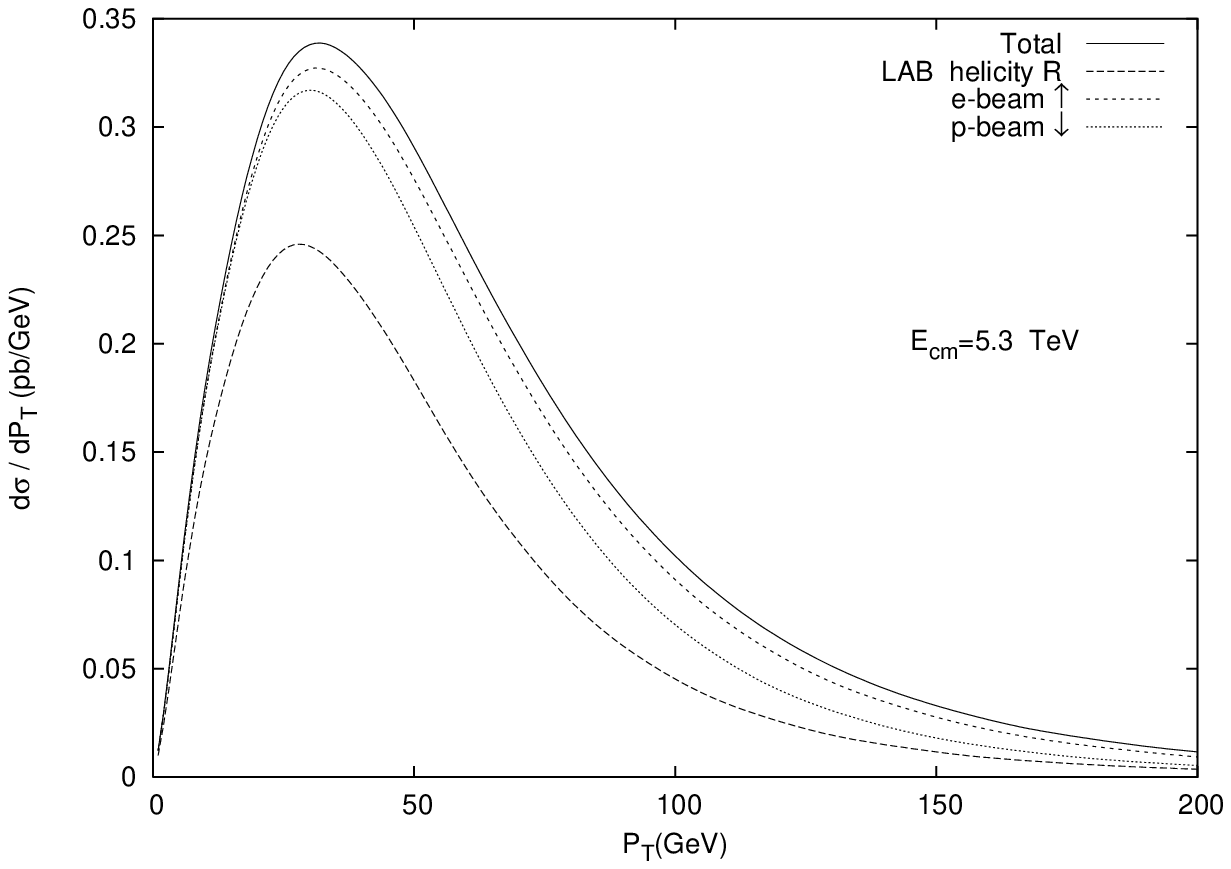}
\caption{Transverse momentum $P_{T}$ distributions
of the singly produced top quarks via $Wg$ fusion at
CLIC+LHC enegy $\sqrt{s}=5.3$ TeV. Dominant spin bases
LAB helicity right, e-beam up, antineutrino up
and unpolarized(Total) case are drawn. Combination of
$2\to 2$  and $2\to 3$ processes are considered
\label{fig2}}
\end{figure}

It is useful to plot the top quark $P_{T}$ distributions to
compare different spin bases. Contributions from
dominant spin components in
LAB helicity right, e-beam up,  p-beam down
states and unpolarized
case(Total) are shown in Fig.~\ref{fig1} for TESLA+HERAp energy
and Fig.~\ref{fig2} for CLIC+LHC energy, using the combination
of $2\to 2$  and $2\to 3$ processes. Similar features in tables
are reflected  in figures too.

In conclusion, we have shown that the single 
top quarks produced in the
$Wg$ fusion channel in $ep$ collision at energies
$\sqrt{s}=1.6$ and 5.3 TeV are used to observe their
high degree spin polarization along the direction
of positron beam  in the
top quark rest frame. The TESLA+HERAp gives better
results leading higher  spin polarization than
CLIC+LHC. Although  $ep$ and $pp$ colliders
are comparable in terms of high spin polarization,
the $Wg$ fusion channel in $ep$ collision is expected
to be much less complicated theoretically
and experimentally than the case in $pp$ collision.

It is anticipated that next-to-leading order corrections
to $Wg$ fusion will improve the results given in
Table~\ref{tab1} and Table~\ref{tab2}.

The experimental conditions to observe the
spin induced angular correlations
with cuts and detector environments still need
to be discussed before a firm  decision.


\begin{thebibliography}{99}
\bibitem{zerwas} I.Bigi, Y. Dokshitzer, V. Khoze, 
J. K\"{u}hn and P. Zerwas, Phys. Lett. {\bf B181} (1986)157.  
\bibitem{mahlon} G. Mahlon and S. Parke, Phys. Lett. 
{\bf B476} (2000)323;
\bibitem{parke}
T.M.P. Tait, Phys. Rev. {\bf D61} (2000)034001;
A.S. Belyaev, E.E. Boos and L.V. Dudko
Phys. Rev. {\bf D59} (1999)075001;
T. Stelzer, Z. Sullivan and
S. Willenbrock, Phys. Rev. {\bf D58} (1998)094021;
G. Mahlon and S. Parke, Phys. Lett.
{\bf B411} (1997)173;
A.P. Heinson, A.S. Belyaev, E.E. Boos,
Phys. Rev. {\bf D56} (1997)3114;
G. Mahlon and S. Parke, Phys. Rev. {\bf D55} (1997)7249;
S. Parke and Y. Shadmi,  Phys. Lett. {\bf B387} (1996)199.
\bibitem{tesla} H. Abramowicz {\it et al.}, 
THERA Collaboration, A contribution to the TESLA 
Technical Design Report, TESLA TDR, Appendices, 
Chapter 2, DESY-01-011FB,2001. 
\bibitem{mahlon2} G. Mahlon and S. Parke, 
Phys. Rev. {\bf D53} (1996)4886.
\bibitem{olness}F. Olness and W.-K. Tung, 
Nucl. Phys. {\bf B308} (1988)813;
R. Barret, H. Haber and D. Soper, 
Nucl. Phys. {\bf B306} (1988)697;
M.A.G. Aivazis, J.C. Collins, F.I. Olness and W.-K. Tung,
Phys. Rev. {\bf D50} (1994)3102.
\bibitem{stelzer} T. Stelzer, Z. Sullivan and 
S. Willenbrock, Phys. Rev. {\bf D56} (1997)5919;
F. Anselmo, B. van Eijk and G. Bordes,
Phys. Rev. {\bf D45} (1992)2312;
G. Bordes and B. van Eijk, Z. Phys. {\bf C57} (1993)81;
G. Bordes and B. van Eijk, Nucl. Phys. {\bf B435}
(1995)23.
\bibitem{stirling} A.D. Martin, R.G. Roberts, W.J. Stirling
and R.S. Thorne, Eur. Phys. J. {\bf C28} (2003)455.
\bibitem{pdg} K. Hagiwara {\it et al.} 
(Particle Data Group),  Phys. Rev. {\bf D66} (2002)010001.
\end{thebibliography}
\end{document}